\documentclass[useAMS,a4paper]{mn2e}
 \usepackage{times}
 \usepackage{epsfig}
 \usepackage{amssymb}

 \title[On the dynamical evolution of 2002 VE$_{68}$] 
       {On the dynamical evolution of 2002 VE$_{68}$} 

 \author[C. de la Fuente Marcos and R. de la Fuente Marcos]
        {C.~de~la~Fuente~Marcos\thanks{E-mail: nbplanet@fis.ucm.es}
         and
         R. de la Fuente Marcos \\
         Universidad Complutense de Madrid,
         Ciudad Universitaria, E-28040 Madrid, Spain}
 \date{Accepted 2012 August 14.
       Received 2012 August  9;
       in original form 2012 July 25}
 \pubyear{2012}
 
\begin{document}
  \maketitle

  \begin{abstract}
     Minor planet 2002 VE$_{68}$ was identified as a quasi-satellite of 
     Venus shortly after its discovery. At that time its data-arc span 
     was only 24 days, now it is 2,947 days. Here we revisit the topic 
     of the dynamical status of this remarkable object as well as look 
     into its dynamical past and explore its future orbital evolution 
     which is driven by close encounters with both the Earth-Moon system 
     and Mercury. In our calculations we use a Hermite integration 
     scheme, the most updated ephemerides and include the perturbations 
     by the eight major planets, the Moon and the three largest 
     asteroids. We confirm that 2002 VE$_{68}$ currently is a 
     quasi-satellite of Venus and it has remained as such for at least 
     7,000 yr after a close fly-by with the Earth. Prior to that 
     encounter the object may have already been co-orbital with Venus or 
     moving in a classical, non-resonant Near-Earth Object (NEO) orbit. 
     The object drifted into the quasi-satellite phase from an L$_4$ 
     Trojan state. We also confirm that, at aphelion, dangerously close 
     encounters with the Earth (under 0.002 AU, well inside the Hill 
     sphere) are possible. We find that 2002 VE$_{68}$ will remain as a 
     quasi-satellite of Venus for about 500 yr more and its dynamical 
     evolution is controlled not only by the Earth, with a 
     non-negligible contribution from the Moon, but by Mercury as well. 
     2002 VE$_{68}$ exhibits resonant (or near resonant) behaviour with 
     Mercury, Venus and the Earth. Our calculations indicate that an 
     actual collision with the Earth during the next 10,000 yr is highly 
     unlikely but encounters as close as 0.04 AU occur with a 
     periodicity of 8 years.  
  \end{abstract}

  \begin{keywords}
     celestial mechanics -- planets and satellites: individual: Venus -- 
     asteroids: individual: 2002 VE68 --
     Solar System: general -- minor planets, asteroids 
  \end{keywords}

  \section{Introduction}
     Minor planet 2002 VE$_{68}$ was discovered by Brian A. Skiff working for the LONEOS Survey at Lowell 
     Observatory on November 11, 2002 and confirmed by the Eschenberg Observatory the following night 
     (Griesser, Skiff \& Spahr 2002)\footnote{http://www.minorplanetcenter.org/mpec/k02/k02V52.html}. With a 
     value of the semi-major axis $a$ = 0.7237 AU very close to that of Venus (0.7233 AU), this Aten asteroid 
     is a Near-Earth Object (NEO) moving in a quite eccentric orbit, $e$ = 0.4104, that makes it a Mercury 
     grazer, Venus crosser and Earth crosser. It has been designated a Potentially Hazardous Asteroid (PHA) by 
     the Minor Planets Center (MPC) and as such has been the target of Doppler studies at Goldstone (Ostro \& 
     Giorgini 2004; Benner et al. 2008; Gavrik \& Gavrik 2008) on November 2002 and 2010. These radar 
     observations suggest that its near-surface is extremely rough. 
     \hfil\par
     A preliminary rotational period of 13.5 h with a light curve amplitude $>$ 0.8 mag (indicating a very 
     elongated body) and an estimated size of 260 m were found by Pravec, Wolf \& Sarounova (2010)\footnote{http://www.asu.cas.cz/$~$ppravec/neo.htm}. 
     Bessel $BVRI$ photometry (Barajas et al. 2011)\footnote{http://www.astronomerstelegram.org/?read=3073} 
     has showed that 2002 VE$_{68}$'s mean colors are compatible with those of an X-type asteroid, perhaps 
     similar to the E-type asteroid 2867 Steins (but also 1114 Lorraine, 5294 Onnetoh, 796 Sarita, 107 Camilla 
     or 3686 Antoku). Barajas et al. (2011) also calculated a synodic period of 13.5 h (confirming the 
     previous preliminary value), an albedo of about 0.25 and an absolute visual magnitude of 20.59 that gives 
     an effective diameter of about 200 m (also consistent to preliminary determinations). With an amplitude 
     of 0.9 mag, its light curve suggests that it may be a contact binary in which two rubble piles orbit a 
     centre of mass in contact with each other (the full details of this research are available from the CURE 
     at LACC web site\footnote{http://www.lacitycollege.edu/academic/departments/physics/cure/reports/ BarajasT\_Sp2011\_Report.pdf}). 
     This physical characterization is consistent with the battered surface suggested by radar data.
     \hfil\par
     Numerical computations by Mikkola et al. (2004) soon revealed that 2002 VE$_{68}$ is moving in a 1:1 mean 
     motion resonance with Venus; more specifically, the asteroid is a quasi-satellite of Venus. As such, 2002 
     VE$_{68}$ is not a real, gravitationally bound satellite but from Venus point of view, the object appears 
     to travel around it over the course of a Venusian year although it actually orbits the Sun. Venus has no 
     known satellites: Sheppard \& Trujillo (2009) completed a survey in search for satellites but no actual 
     moons down to about 0.3 km in radius were detected. At the time of its identification as quasi-satellite 
     of Venus, the arc length of 2002 VE$_{68}$ was only 24 days so that its orbit was not yet well known. The 
     orbit has been improved significantly over the years and now it has an arc length of 2,947 days; besides, 
     2002 VE$_{68}$ has also been observed by radar (Ostro \& Giorgini 2004; Benner et al. 2008). Here we 
     revisit the topic of the current dynamical status of this remarkable object as well as look into its 
     dynamical past and explore its future orbital evolution which is driven by close encounters with both the 
     Earth and Mercury. In our calculations we use the most updated ephemerides and include the perturbations 
     by the eight major planets. In addition, we include perturbations from the Moon and the three largest 
     asteroids, (1) Ceres, (2) Pallas and (4) Vesta. 
     \hfil\par
     This paper is organized as follows: in Section 2 the details of our numerical integrations are given. In 
     Section 3, we present the results of our simulations. We discuss our results in Section 4. In Section 5 
     we compare our results with those obtained by Mikkola et al. (2004) and our conclusions are summarized in 
     Section 6. 

  \section{Simulations}
     In our calculations we directly integrate the full equations of motion using the Hermite scheme described 
     by Makino (1991) and Aarseth (2003). Additional simulations were completed using the time-symmetric 
     Hermite method described by Kokubo, Yoshinaga \& Makino (1998) but it was found that, for the problem 
     studied here, its overall performance was lower and the results largely identical. The Hermite scheme 
     allows efficient numerical integration of the entire solar system thanks to the use of a block-step 
     scheme (Aarseth 2003) in which suitably quantized time-steps enable following the orbits of Mercury or 
     planetary satellites and trans-Neptunian objects simultaneously. The standard versions of these scalar 
     $N$-body codes are publicly available from the IoA web site\footnote{http://www.ast.cam.ac.uk/$\sim$sverre/web/pages/nbody.htm}. 
     These versions have been modified in order to study the orbital evolution of 2002 VE$_{68}$. 
     \hfil\par
     For accurate initial positions and velocities we used the Heliocentric ecliptic Keplerian elements 
     provided by the JPL\footnote{http://ssd.jpl.nasa.gov/sbdb.cgi} and initial positions and velocities based 
     on the DE405 planetary orbital ephemerides (Standish 1998)\footnote{http://ssd.jpl.nasa.gov/?planet\_pos} 
     referred to the barycentre of the Solar System. Orbits are calculated forward and backward in time. Our 
     reference calculations include the perturbations by eight major planets (Mercury to Neptune) and treat 
     the Earth and the Moon as a single object. As a zeroth order approximation, the Earth and the Moon can be 
     replaced with a ficticious body at their barycentre but, in relative terms, the Moon is the largest 
     satellite in the Solar System and, among satellites, it has one of the largest semi-major axes. 
     Therefore, a better approximation is to resolve the Earth and the Moon as two separate bodies and assume 
     that they are point mass objects and the only force acting between them is Newtonian gravitation, i.e. 
     tidal dissipation is neglected. This will be our second physical model. In all cases we consider point 
     (constant) mass objects orbiting in a conservative system; therefore, relativistic effects are ignored. 
     \hfil\par
%
%
     \begin{figure*}
        \centerline{\hbox{
         \epsfig{figure=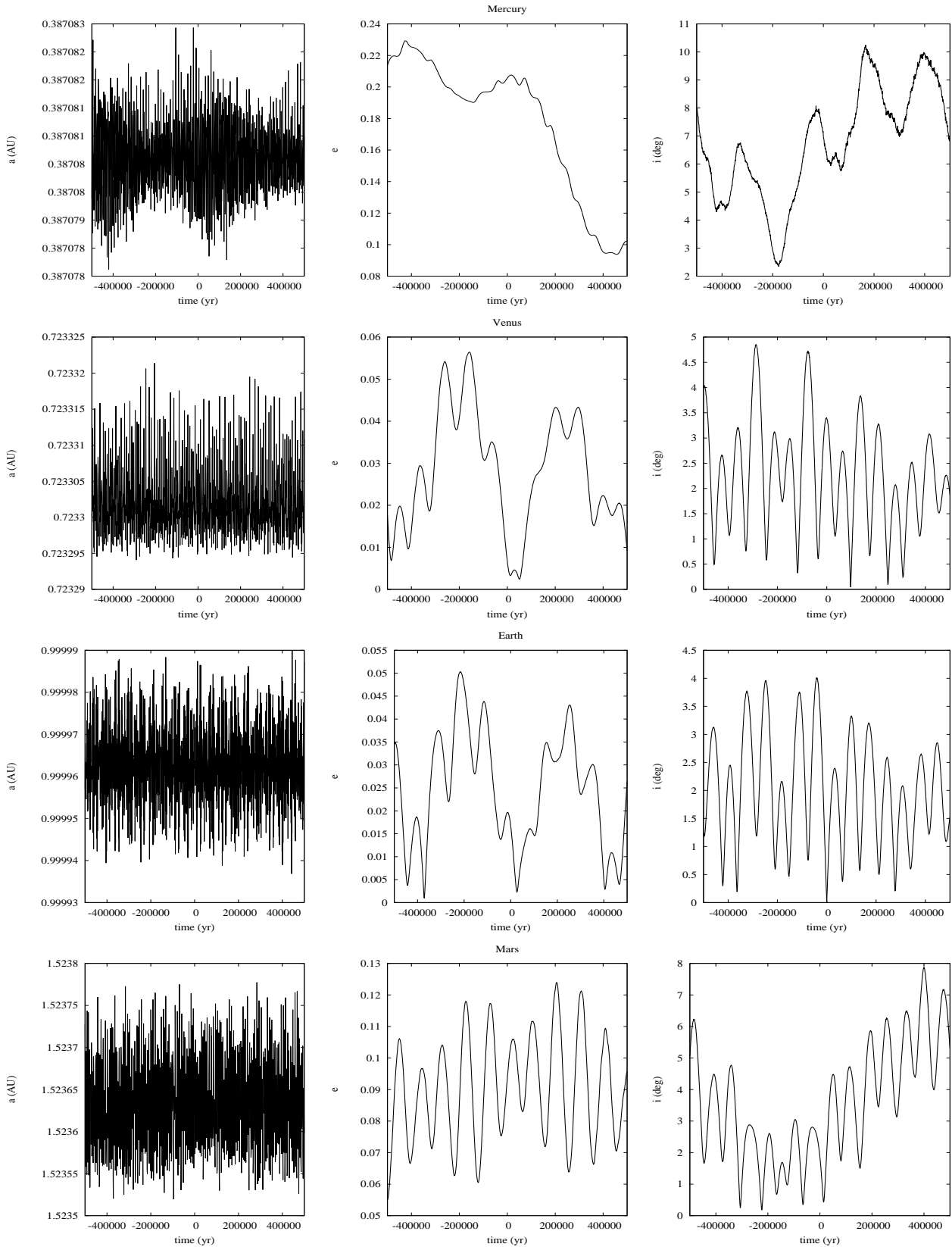,height=22cm,width=17cm,angle=0}
          }}
         \caption{Evolution of the semi-major axes, eccentricities and inclinations of the four rocky planets 
                  (Mercury to Mars) from -0.5 to +0.5 Myr. These variations result from the gravitational
                  perturbations on the motion of each planet from all the other planets of the Solar System.   
                 }
         \label{ssevol0}
     \end{figure*}
%
%
%
%
     \begin{figure*}
        \centerline{\hbox{
         \epsfig{figure=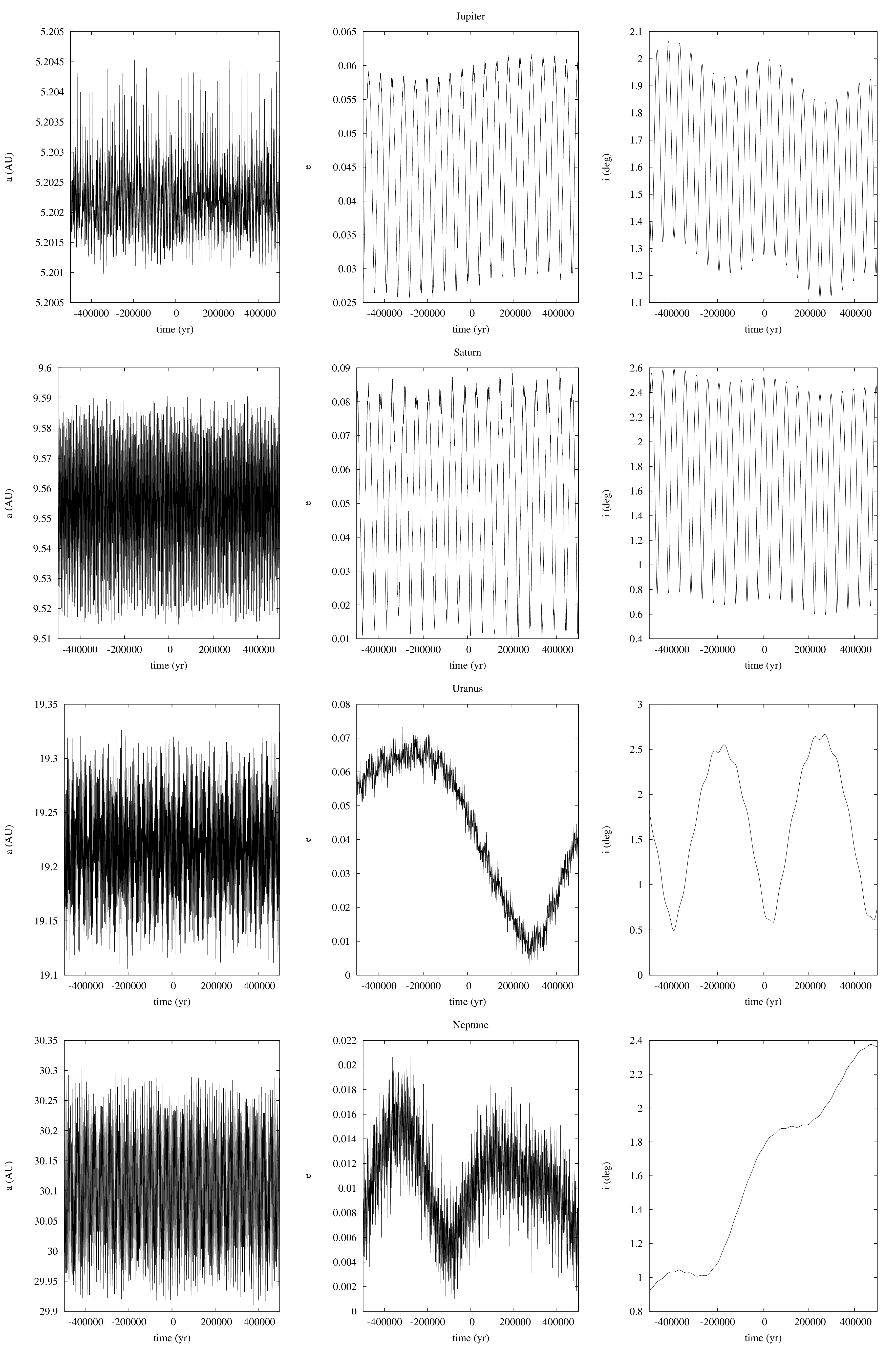,height=22cm,width=17cm,angle=0}
          }}
        \caption{Semi-major axes, eccentricities and inclinations of the four giant planets (Jupiter to 
                 Neptune).   
                }
        \label{ssevol1}
     \end{figure*}
%
%
     To ensure that the code used in this study was appropriate for the task, a significant amount of testing 
     was performed and its numerical integrations have been validated against publicly available results 
     obtained by other authors using other algorithms and physical models. Following Varadi, Runnegar \& Ghil 
     (2003), we also estimated the actual integration errors by computing the same orbits with the same 
     physical model but with two different step sizes (or, more properly, blocks of them). In the 
     predictor-corrector algorithm embedded into the Hermite scheme, the overall "step size" is controlled by 
     an input amount called the time-step convergence parameter for total force polynomials, $\eta$, which is 
     a dimensionless accuracy parameter (Aarseth 2003). For values of $\eta$ in the range 10$^{-5}$-10$^{-7}$, 
     the results (and the integration errors) are similar but the error in the total energy is minimal for 
     $\eta$ = 1$\times10^{-6}$. Then, relative errors in the total energy are as low as 5$\times10^{-15}$ 
     after 0.4 Myr. The relative error of the total angular momentum is several orders of magnitude smaller. 
     In Figs. \ref{ssevol0} and \ref{ssevol1} the evolution of the semi-major axes, eccentricities and 
     inclinations of the eight major planets are shown for 1 Myr centred on the epoch JD2456000.5. The output 
     time-step for these plots is 10$^3$ yr and no filtering or smoothing was applied to the data. This output 
     cadence is not expected to introduce any aliasing.
     \hfil\par
     The evolution of the semi-major axes, eccentricities and inclinations of the inner planets (Mercury to 
     Mars) from -0.5 to +0.5 Myr are plotted in Fig. \ref{ssevol0}. These results are similar to those in
     Laskar (1988, 1990), Quinn, Tremaine \& Duncan (1991), Laskar, Joutel \& Boudin (1993) or Ito \& Tanikawa
     (2002). Major differences appear for Mercury but only prior to -250,000 yr and after 400,000 yr. The 
     impact of these deviations on our results is expected to be negligible. The corresponding orbital 
     evolution for the outer planets (Jupiter to Neptune) is displayed in Fig. \ref{ssevol1}. Eccentricities 
     in de Pater \& Lissauer (2010), Fig. 2.14, match our results very well.  
     \hfil\par
\begin{figure}
        \centerline{\hbox{
         \epsfig{figure=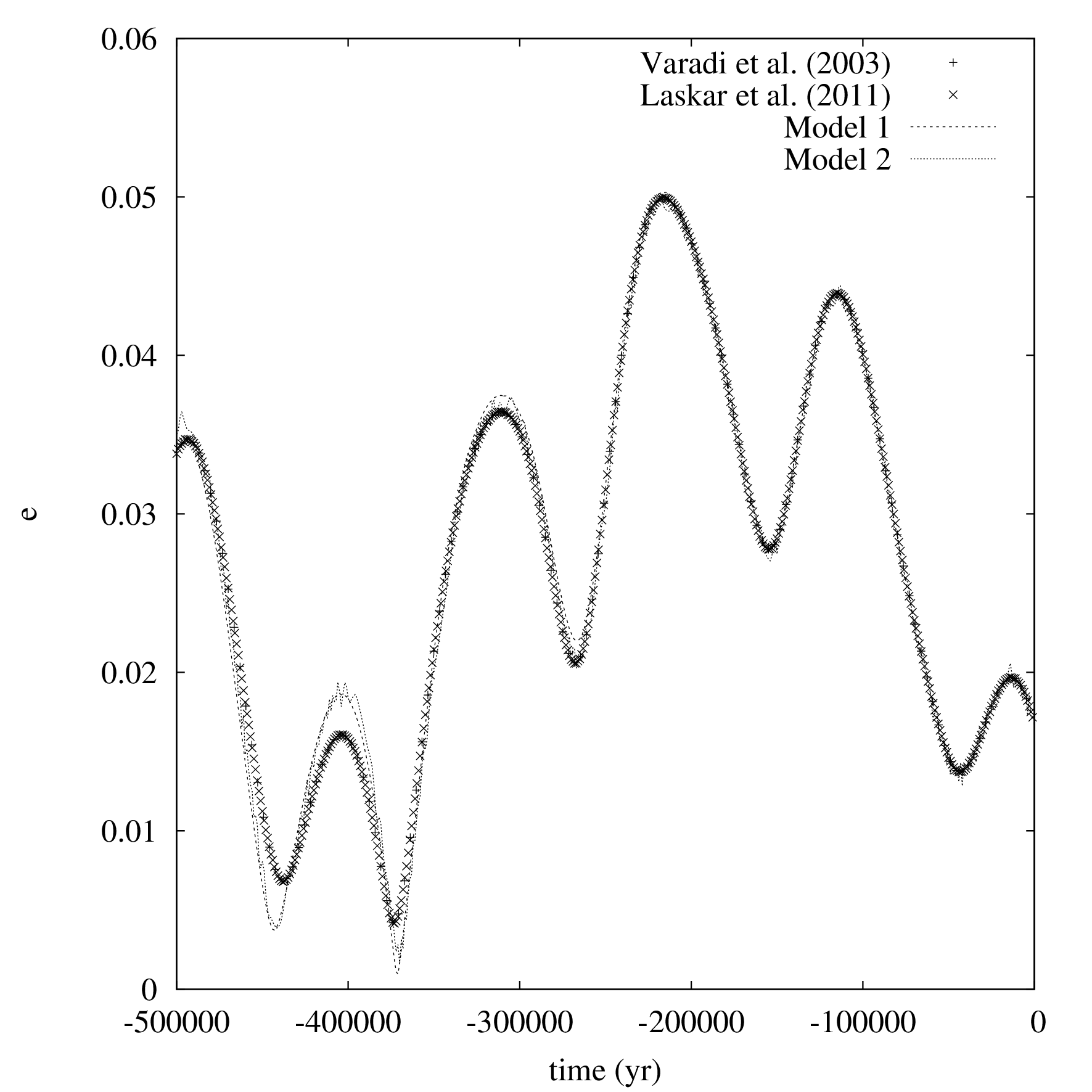,height=6cm,width=8cm,angle=0}
          }}
        \caption{Evolution of Earth's orbital eccentricity according to Varadi et al. (2003) and Laskar et al. 
                 (2011) and a detail of our validation simulations (see Fig. \ref{ssevol0}, Earth). The time 
                 axis indicates that only data from the past are being displayed. The model displayed from 
                 Laskar et al. (2011) is Model A. Model 1 includes the eight major planets (Mercury to 
                 Neptune) and treat the Earth and the Moon as a single object. In Model 2, the Earth-Moon system 
                 is resolved as two separate bodies. Differences become noticeable after 250,000 years. This 
                 figure does not show large differences between our results and those of Varadi et al. (2003) 
                 or Laskar et al. (2011).
                }
        \label{validation}
     \end{figure}

     We have carried out a more detailed comparison between our results and those from Varadi et al. (2003) 
     and Laskar et al. (2011). These authors have performed long-term numerical simulations of the orbits of 
     the major planets in our Solar System using a variety of models and integration algorithms. It is obvious
     that the precision of our present astronomical computations within the simulated time frame is comparable 
     to that in these recent studies as seen in Fig. \ref{validation}. The figure does not show any large 
     differences between our results and those of Varadi et al. (2003) or Laskar et al. (2011). There are no 
     unusual features that would hint at a major problem with our models or integration method. Here, the 
     results of Varadi et al. (2003) paper have been obtained from Prof. Varadi's web site\footnote{http://www.astrobiology.ucla.edu/OTHER/SSO/}. 
     The data of Laskar et al. (2011) have been downloaded from the Astronomical Solutions for Earth 
     paleoclimates web site\footnote{http://www.imcce.fr/Equipes/ASD/insola/earth/La2010/index.html}. We 
     interpret these positive comparisons as an explicit validation of our calculations.
%
%
%
%

  \section{Results}
     The Venus quasi-satellite 2002 VE$_{68}$ is an unusual object that is directly perturbed by three of the 
     inner planets, Mercury, Venus and the Earth. The object has a very significant eccentricity (0.41) and it 
     is an obvious candidate to be in an unstable orbit. The overall stability of asteroids in co-orbital 
     motion with Venus has been studied multiple times (e.g. Mikkola \& Innanen 1992; Tabachnik \& Evans 2000; 
     Scholl, Marzari \& Tricarico 2005; Morais \& Morbidelli 2006). Results from these studies indicate that 
     any hypothetical primordial population of Venusian co-orbitals could not possibly have survived until the 
     present time and that any current population of co-orbital objects must be transient in nature. Asteroids
     in co-orbital motion with Venus undergo multiple captures/ejections in/from the 1:1 mean motion resonance 
     and the average duration of one of these events (capture to ejection) is 32,000 yr (Morais \& Morbidelli 
     2006). During these episodes they may become quasi-satellites or librate on horseshoe orbits or tadpole 
     orbits (Scholl et al. 2005). 
     \hfil\par 
     Here we present the results for the nominal orbit in Table \ref{elements}. In addition to these 
     calculations using the nominal orbital elements and for Model 3 (see below), we have performed 100 
     control simulations using sets of orbital elements derived from the nominal ones using the uncertainties 
     in Table \ref{elements}, at 3-$\sigma$. These control integrations take into account the uncertainties in 
     observation and orbital determination. As listed in Table \ref{elements}, the errors are very small and 
     the results for the control orbits are very close to those of the nominal orbit shown in Figs. \ref{qs} 
     and \ref{mld}. Mikkola et al. (2004) already pointed out that the orbit of 2002 VE$_{68}$ is quite 
     chaotic. Chaotic orbits are not only sensitive to changes in the initial conditions but also to different 
     dynamical models. Starting from the initial conditions described above, we integrate the orbits up to 
     20,000 yr in both directions of time although only the time interval (-10,000, 10,000) yr will be 
     displayed in our figures. The output time-step (time resolution) in all the figures is 0.01 yr (3.65 
     days) and no filtering or smoothing has been applied to the data. Again, this output frequency is not 
     expected to introduce any aliasing in our results. Our calculations do not include any modeling of the 
     Yarkovsky and YORP effects (see, e.g., Bottke et al. 2006).
%
%
         \begin{table}
          \fontsize{8} {11pt}\selectfont
          \tabcolsep 0.10truecm
          \caption{Heliocentric orbital elements of 2002 VE$_{68}$ used in this research. 
                   (Epoch = JD2456200.5, 2012-Sep-30.0; J2000.0 ecliptic and equinox.)
                   Values include the 1-$\sigma$ uncertainty (Source: JPL Small-Body
                   Database).
                  }
          \begin{tabular}{ccc}
           \hline
            Semi-major axis, $a$                  & = & 0.7236659191$\pm$0.0000000005 AU \\
            Eccentricity, $e$                     & = & 0.41035660$\pm$0.00000005 \\
            Inclination, $i$                      & = & 9.005869$\pm$0.000013 $^{\circ}$ \\
            Longitude of ascending node, $\Omega$ & = & 231.584442$\pm$0.000005 $^{\circ}$ \\
            Argument of perihelion, $\omega$      & = & 355.463207$\pm$0.000014 $^{\circ}$ \\
            Mean anomaly, $M$                     & = & 202.88183$\pm$0.00003 $^{\circ}$ \\ 
           \hline
          \end{tabular}
          \label{elements}
         \end{table}
%
%

     \subsection{2002 VE$_{68}$: current dynamical status}
        2002 VE$_{68}$ was identified as quasi-satellite of Venus by Mikkola et al. (2004). It was the first
        quasi-satellite observed and identified as such. The quasi-satellite dynamical state is a specific 
        configuration of the 1:1 mean motion resonance with a host planet in which the object appears to 
        travel around the planet but is not gravitationally bound to it: the body librates around the 
        longitude of its associated planet but its trajectory is not closed. The term "quasi-satellite" was 
        first used in a scientific paper by Danielsson \& Ip (1972a) in the context of explaining the resonant 
        behaviour of the NEO (1685) Toro with the Earth (Danielsson \& Ip 1972b). However, this early mention 
        was not directly connected with the topic of co-orbital bodies and it is usually considered that the 
        term was first introduced and popularized among the scientific community by Mikkola \& Innanen (1997), 
        although the concept behind it was initially studied by Jackson (1913) and the energy balance 
        associated to the resonant state was first analyzed by H\'enon (1969). Further analysis was carried 
        out by Szebehely (1967), Benest (1976), Dermott \& Murray (1981) and Lidov \& Vashkov'yak (1994a,b). 
        Most of this early work was completed in the context of the restricted elliptic three-body problem. 
        \hfil\par
%
%
     \begin{figure}
        \centerline{\hbox{
         \epsfig{figure=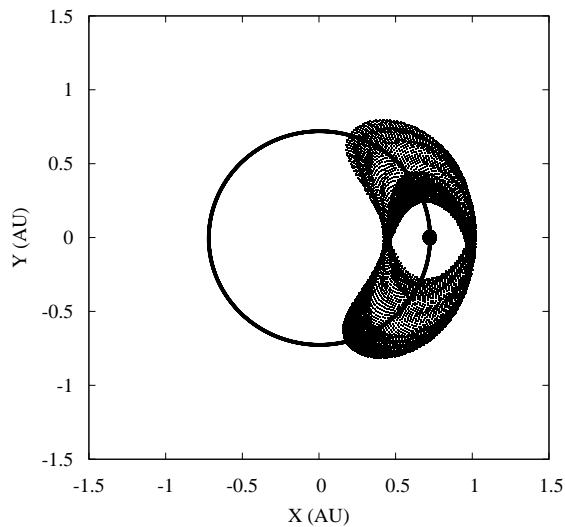,height=9cm,width=9cm,angle=0}
          }}
        \caption{The motion of 2002 VE$_{68}$ for the next 150 yr projected onto the ecliptic plane. The 
                 coordinate system rotates with Venus. The orbit of Venus is also plotted and the actual 
                 position of Venus indicated. These results correspond to Model 3 (see the text for details). 
                 The quasi-satellite appears to follow a precessing kidney-shaped retrograde path when viewed 
                 from Venus. This figure is equivalent to Fig. 1 in Mikkola et al. (2004). 
                }
        \label{qs}
     \end{figure}
%
%
%
%
     \begin{figure}
        \centerline{\hbox{
         \epsfig{figure=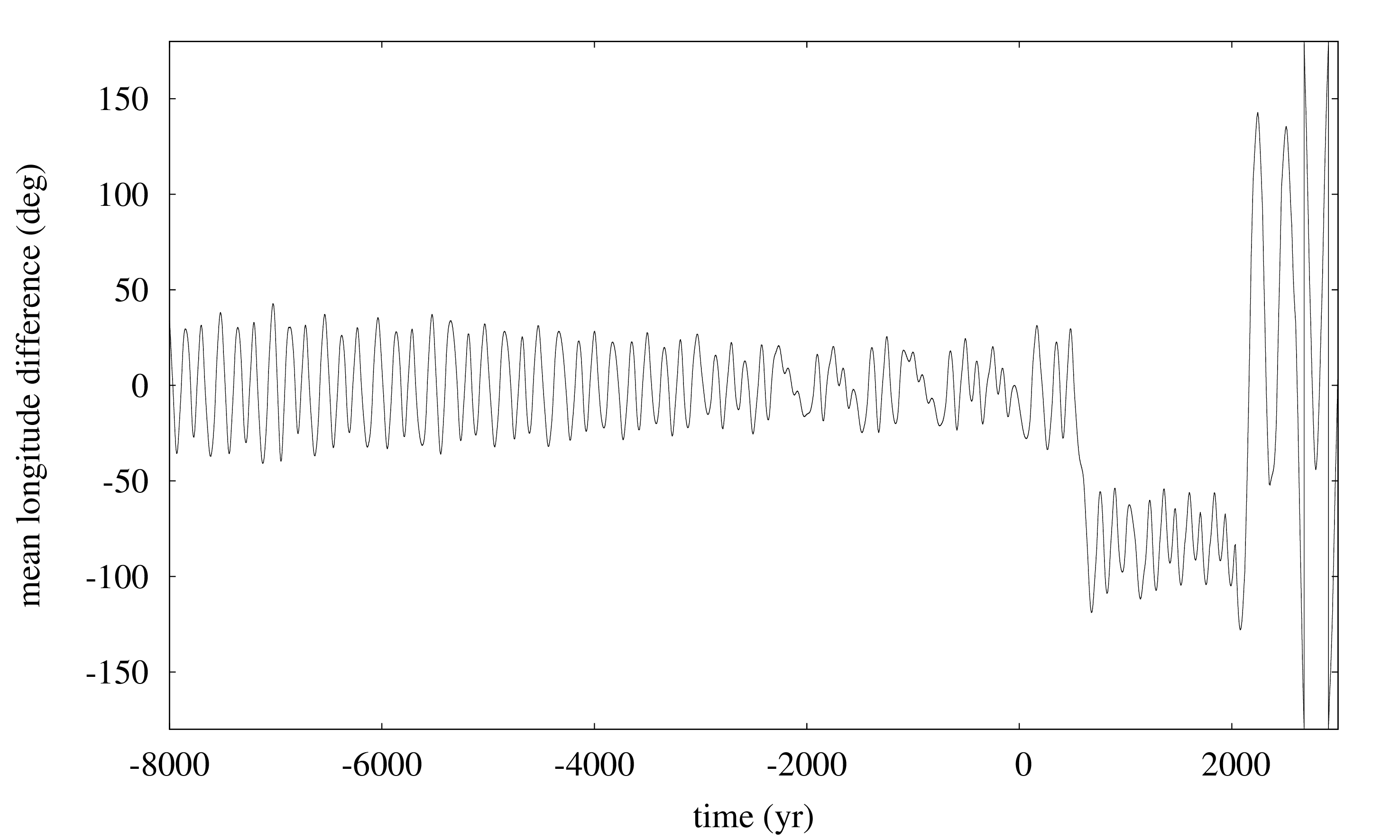,height=6cm,width=8cm,angle=0}
          }}
        \caption{The mean longitude difference of 2002 VE$_{68}$ and Venus. It currently librates around 
                 0$^{\circ}$, that is characteristic of an object in the quasi-satellite dynamical state. 
                 These results correspond to Model 3 (see the text for details). This figure is equivalent to 
                 Fig. 2 in Mikkola et al. (2004).
                }
        \label{mld}
     \end{figure}
%
%
        For an asteroid following a quasi-satellite orbit, the key parameter to study is the difference 
        between the mean longitudes of the asteroid and its host planet or relative mean longitude. The mean 
        longitude of an object is given by $\lambda$ = $M$ + $\Omega$ + $\omega$, $M$ is the mean anomaly, 
        $\Omega$ is the longitude of ascending node and $\omega$ is the argument of perihelion. When the 
        relative mean longitude librates around 0$^{\circ}$, the object is in the quasi-satellite dynamical 
        state, if it librates around 60$^{\circ}$, the objects is called an $L_4$ Trojan, when it librates 
        around -60$^{\circ}$ (or 300$^{\circ}$), it is an $L_5$ Trojan, if the libration amplitude is larger 
        than 180$^{\circ}$, it is said that the object follows a horseshoe orbit and when the relative mean 
        longitude circulates (does not oscillate around a certain value or oscillates freely) we say that the 
        object is no longer in a 1:1 mean motion resonance with the planet, i.e. it becomes a passing object. 
        \hfil\par
        In Fig. \ref{qs} we plot the motion of 2002 VE$_{68}$ for the next 150 yr in a coordinate system that 
        rotates with Venus. It appears to follow a precessing kidney-shaped retrograde path when viewed from 
        Venus over the course of a Venusian year. This result is equivalent to that in Fig. 1 in Mikkola et 
        al. (2004). In Fig. \ref{mld} we plot the mean longitude difference of 2002 VE$_{68}$ and Venus: its 
        value currently librates around 0$^{\circ}$, i.e. the object oscillates around the mean longitude of 
        Venus. This figure is equivalent to Fig. 2 in Mikkola et al. (2004). In both Figs. \ref{qs} and 
        \ref{mld}, data from Model 3 are plotted (see details below). We confirm that the minor planet 2002 
        VE$_{68}$ is a quasi-satellite of Venus, it has remained in its present orbit for thousands of years 
        and will continue trapped in this resonant state for another 500 years.
     
     \subsection{Model 1}
\begin{figure}
        \centerline{\hbox{
         \epsfig{figure=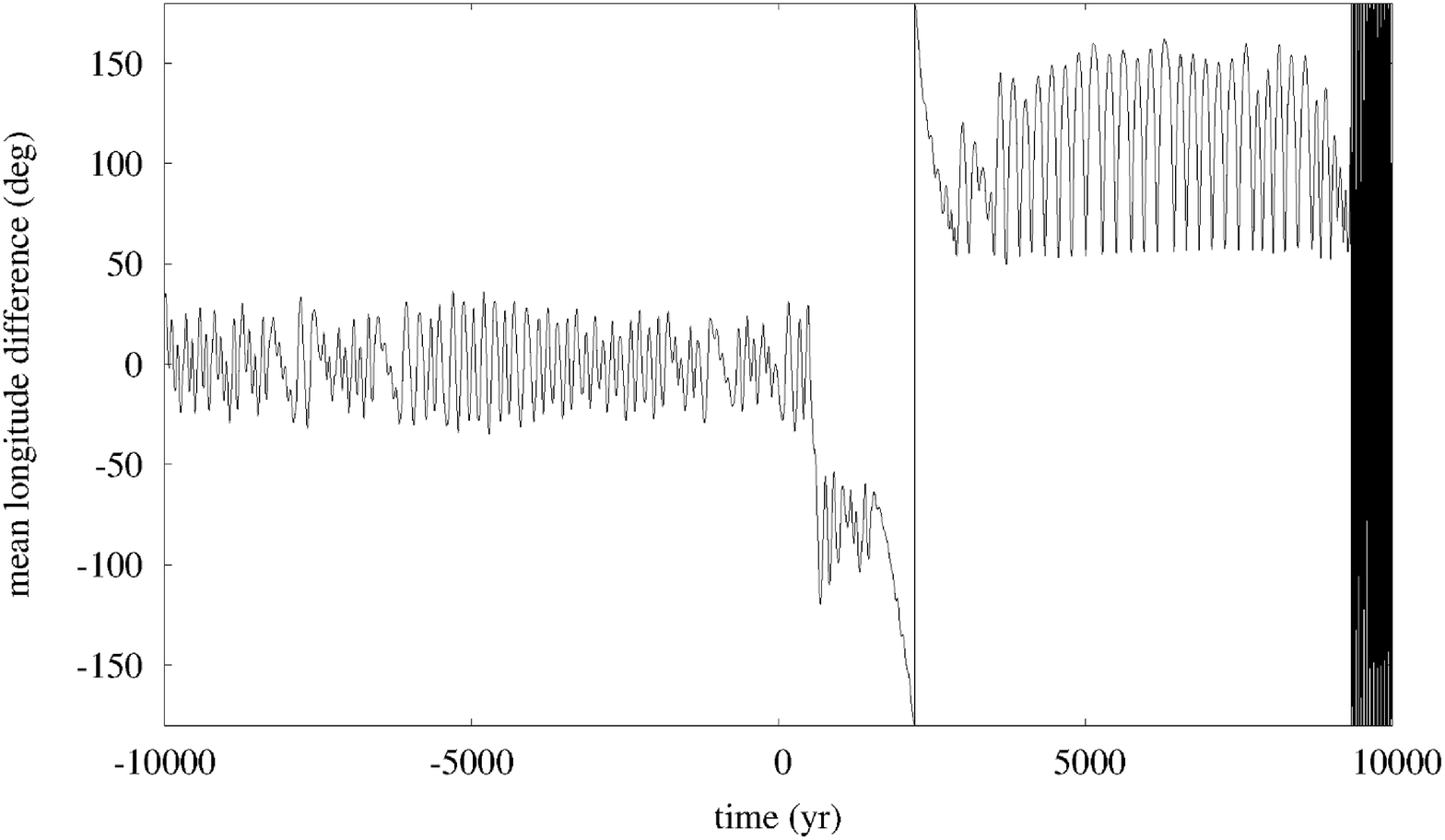,height=6cm,width=8cm,angle=0}
          }}
        \caption{The mean longitude difference of 2002 VE$_{68}$ and Venus in the time interval (-10,000, 
                 10,000) yr from Model 1. The mean longitude difference currently librates around 0$^{\circ}$.
                 It will leave the quasi-satellite dynamical state in 500 yr from now. 
                }
        \label{mld1}
     \end{figure}

        Here we show the results from our Model 1 that includes the perturbations by the eight major planets 
        on 2002 VE$_{68}$ for the nominal orbit in Table \ref{elements}. The mean longitude difference of 2002 
        VE$_{68}$ and Venus is displayed in Fig. \ref{mld1}. As pointed out above, the relative mean longitude 
        currently librates around 0$^{\circ}$. 2002 VE$_{68}$ has remained for some time in the 1:1 mean 
        motion resonance with Venus alternating among the various resonant states: $L_{5}$ Trojan, $L_{4}$ 
        Trojan (tadpole orbits), horseshoe orbit, quasi-satellite or combinations of two or more of these 
        orbits (some of them not displayed). 
        \hfil\par
%
%
\begin{figure*}
        \centerline{\hbox{
         \epsfig{figure=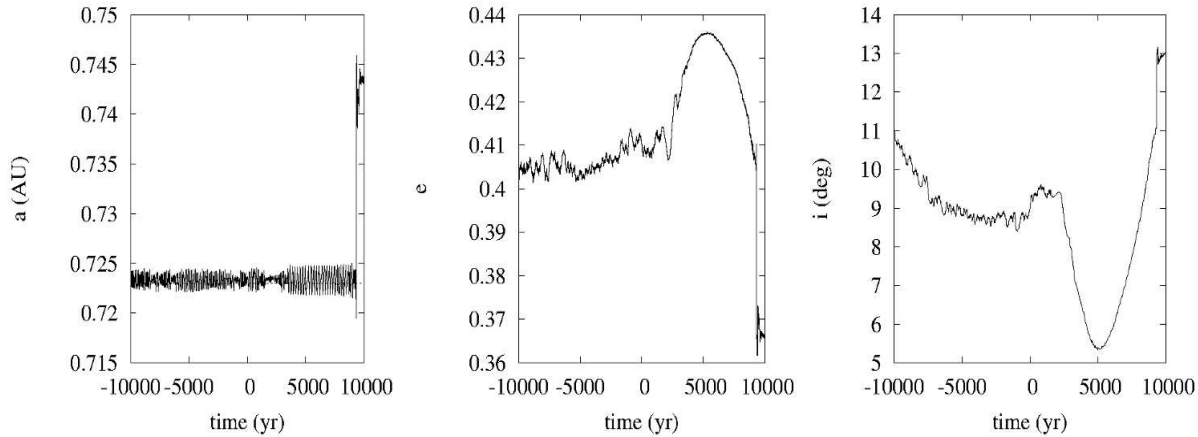,height=6cm,width=16cm,angle=0}
          }}
        \caption{Orbital elements of 2002 VE$_{68}$ from Model 1. The value of the semi-major axis of Venus is
                 also plotted (left panel). When the object is in the quasi-satellite dynamical state, the 
                 eccentricity increases (central panel), the inclination decreases (right panel) and the value
                 of the semi-major axis remains almost constant (left panel), oscillating around the value of 
                 the semi-major axis of Venus.
                }
        \label{orbele1}
     \end{figure*}

%
%
        As expected for an object moving in a rather eccentric orbit and submitted to the direct perturbation 
        of multiple planets, its orbit is certainly unstable. The orbital elements of 2002 VE$_{68}$ are 
        plotted in Fig. \ref{orbele1}. When the object enters the quasi-satellite dynamical state, the 
        eccentricity increases, the inclination decreases and the value of the semi-major axis remains almost 
        constant oscillating around the value of the semi-major axis of Venus. Large variations in both 
        eccentricity and inclination are observed when the object goes from the $L_{5}$ Lagrangian point to 
        $L_{4}$. 
        \hfil\par
%
%
%
%
%
%
%
     \begin{figure}
        \centerline{\hbox{
         \epsfig{figure=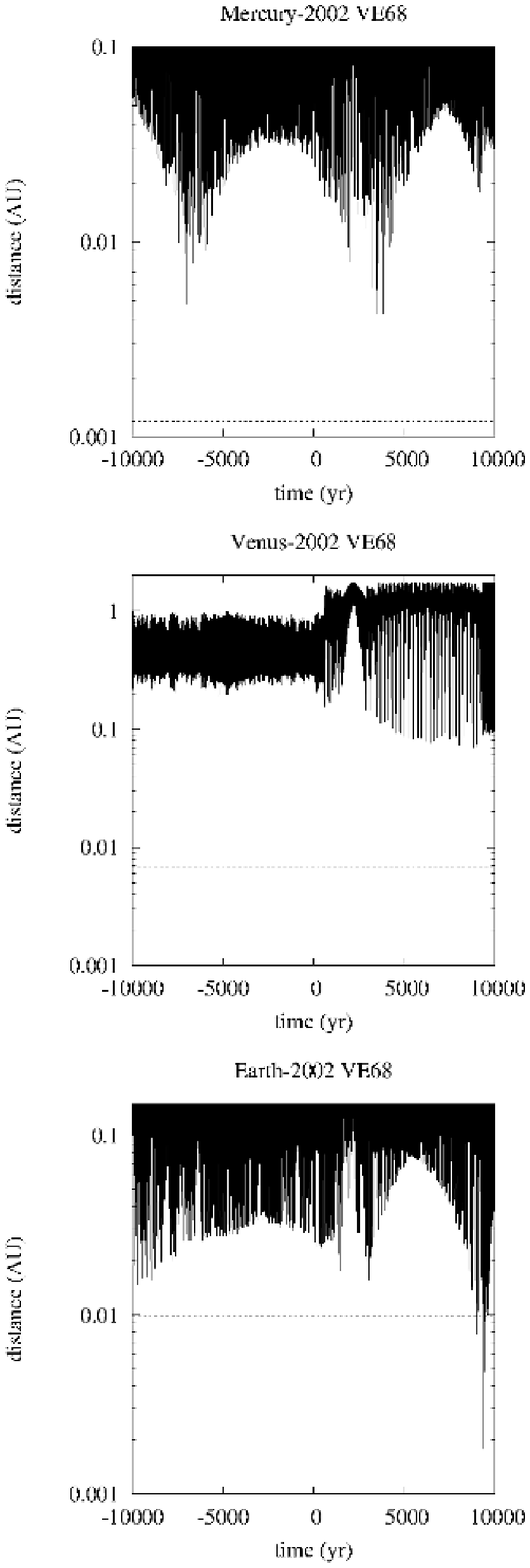,height=15cm,width=8cm,angle=0}
          }}
        \caption{The distance of 2002 VE$_{68}$ from Mercury (top panel), Venus (middle panel) and the Earth 
                 (bottom panel) from Model 1. The value of the Hill sphere radius (see the text) for each 
                 planet is also displayed. The middle panel is equivalent to Fig. 3 in Mikkola et al. (2004).
                 The bottom panel is equivalent to Fig. 4 in Mikkola et al. (2004).
                }
        \label{distance1}
     \end{figure}
%
%
        Transitions between the various resonant phases are triggered by close encounters with planets: the 
        object is a Mercury grazer, Venus crosser and Earth crosser. In Fig. \ref{distance1} we show that the 
        distance of 2002 VE$_{68}$ from Venus remains larger than 0.1 AU until the object is ejected from the 
        quasi-satellite dynamical state into the $L_{5}$ Lagrangian point to become a Venus Trojan about 500 
        yr from now. Encounters with Venus do not appear to be the cause of 2002 VE$_{68}$ entering or leaving 
        the quasi-satellite phase. However, in the cases of both Mercury and the Earth, the distance of 
        closest approach becomes relatively (Mercury) or dangerously close (Earth) to the value of the Hill 
        sphere radius for the respective planet. The Hill sphere radius, $r_H$, is the limiting radius for 
        orbits of planetary satellites in the presence of the Sun's gravitational field and it is given by: 
        $r_H \approx a (1 - e) (m/(3 M_{\odot}))^{1/3}$, where $a$ is the semi-major axis of the planet, $e$ 
        is the eccentricity of its orbit, $m$ is the mass of the planet and $M_{\odot}$ is the mass of the Sun 
        (Hamilton \& Burns 1992). For Mercury, the Hill radius is 0.0012 AU and the closest approach is nearly 
        0.004 AU (although encounters as close as 0.003 AU are observed in our calculations) but for the 
        Earth, the Hill radius is 0.0098 AU and the closest approaches are at 0.002 AU (0.0018 AU, 9,300 yr 
        from now). 
        \hfil\par
        The object is injected into the quasi-satellite dynamical state after a close encounter with the Earth 
        at about 0.007 AU, 11,000 yr ago (not shown in Fig. \ref{distance1}). In contrast, the closest 
        aproaches with Venus are at 0.07 AU but its Hill radius is 0.0067 AU. It is clear, that the current 
        resonant behaviour of 2002 VE$_{68}$ with Venus is mainly controlled by the Earth but the secondary 
        role of Mercury can not be neglected (the closest approach is just 2.5 times outside the Hill sphere 
        of Mercury). At this point, let us remind the reader that in Model 1 we consider the Earth-Moon system 
        as a single object and we follow its barycentre; therefore, when we consider the distance from 2002 
        VE$_{68}$ to the Earth we actually mean Earth's barycentre. The average Earth-Moon separation is about 
        0.0025 AU; close encounters can, in principle, make 2002 VE$_{68}$ pass between the Earth and the Moon 
        in the future; therefore, we must conclude that our natural satellite will play a non-negligible role 
        in the outcome of those close encounters and a better physical model is required in order to obtain 
        more reliable results. On the other hand, the dynamical role of Venus may become more important after 
        the object is ejected from the quasi-satellite state (about 500 yr from now) and eventually evolves 
        into a passing object (10,000 yr from now): then, approaches close to the Hill sphere of Venus are 
        possible (0.0075 AU, nearly 19,500 yr from now). 

     \subsection{Model 2}
        Our previous calculations clearly indicate that, in the case of 2002 VE$_{68}$, encounters with the 
        Earth as close as 0.002 AU are possible. This is 0.8 times the distance between the Earth and the 
        Moon. Almost certainly, the dynamical role of our natural satellite on the outcome of these close 
        encounters cannot be neglected. In order to further investigate this claim, we will repeat the 
        calculations including the perturbations by the eight major planets and the Moon on 2002 VE$_{68}$ for 
        the nominal orbit in Table \ref{elements}. Now, the Earth-Moon system is resolved as two separate 
        bodies and assumed to be made of point mass objects; the only force acting between them is Newtonian 
        gravitation, i.e. tidal dissipation is neglected. 
        \hfil\par
%
%
     \begin{figure}
        \centerline{\hbox{
         \epsfig{figure=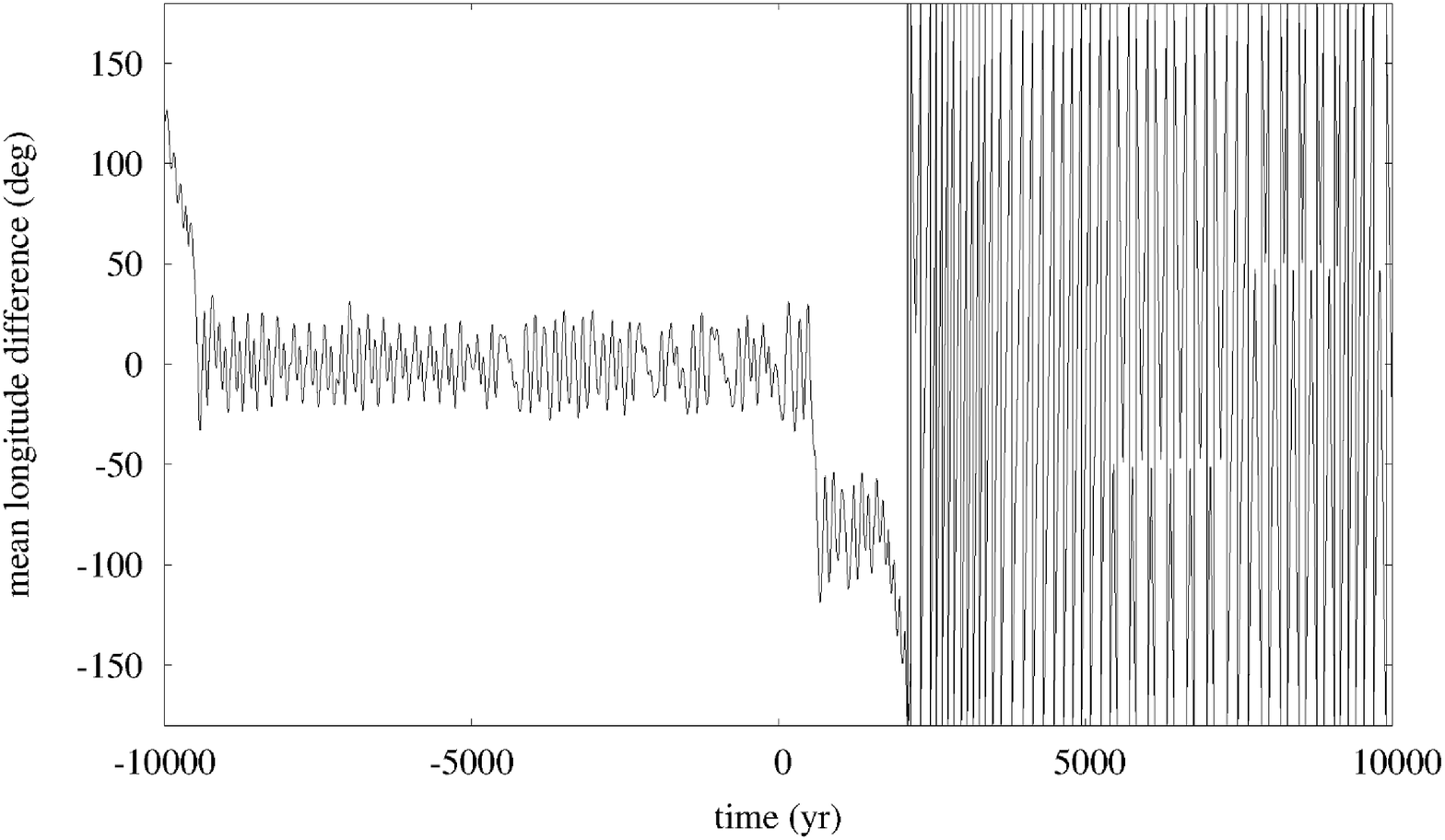,height=6cm,width=8cm,angle=0}
          }}
        \caption{The mean longitude difference of 2002 VE$_{68}$ and Venus in the time interval (-10,000, 
                 10,000) yr from Model 2. In the time interval -7,000 to +1,000 yr from now the evolution of 
                 the mean longitude difference is very similar to that from Model 1 (see Fig. \ref{mld1}) but, 
                 after leaving the quasi-satellite dynamical state, the asteroid does not become a long-term 
                 Venus Trojan. 
                }
        \label{mld2}
     \end{figure}
\begin{figure*}
        \centerline{\hbox{
         \epsfig{figure=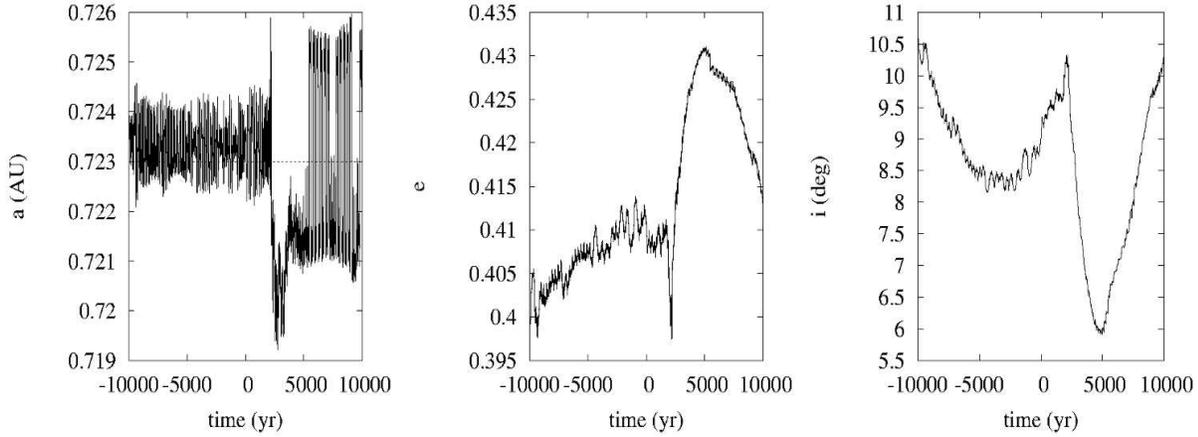,height=6cm,width=16cm,angle=0}
          }}
        \caption{Orbital elements of 2002 VE$_{68}$ from Model 2.
                }
        \label{orbele2}
     \end{figure*}

%
%
        Figure \ref{mld2} confirms that 2002 VE$_{68}$ has been a quasi-satellite of Venus for some time 
        although in these calculations it entered the quasi-satellite phase about 7,000 yr ago not 11,000 yr 
        ago like in Model 1. In the time interval -7,000 to +1,000 yr from now the evolution of the mean 
        longitude difference is quite similar to that in Fig. 2 of Mikkola et al. (2004). Now the object, 
        after leaving the L$_5$ Venus Trojan location, quickly enters the horseshoe-like phase. It does not 
        inmediately become an $L_{4}$ Trojan like in Model 1. Consistently, Fig. \ref{orbele2} shows 
        differences with respect to Fig. \ref{orbele1} although they exhibit similar behaviour in the time 
        interval -7,000 to +1,000 yr from now. During the horseshoe-like phase, the semi-major axis remains 
        oscillating (but with larger amplitude) around the value of the semi-major axis of Venus. 
        \hfil\par
%
%
\begin{figure}
        \centerline{\hbox{
         \epsfig{figure=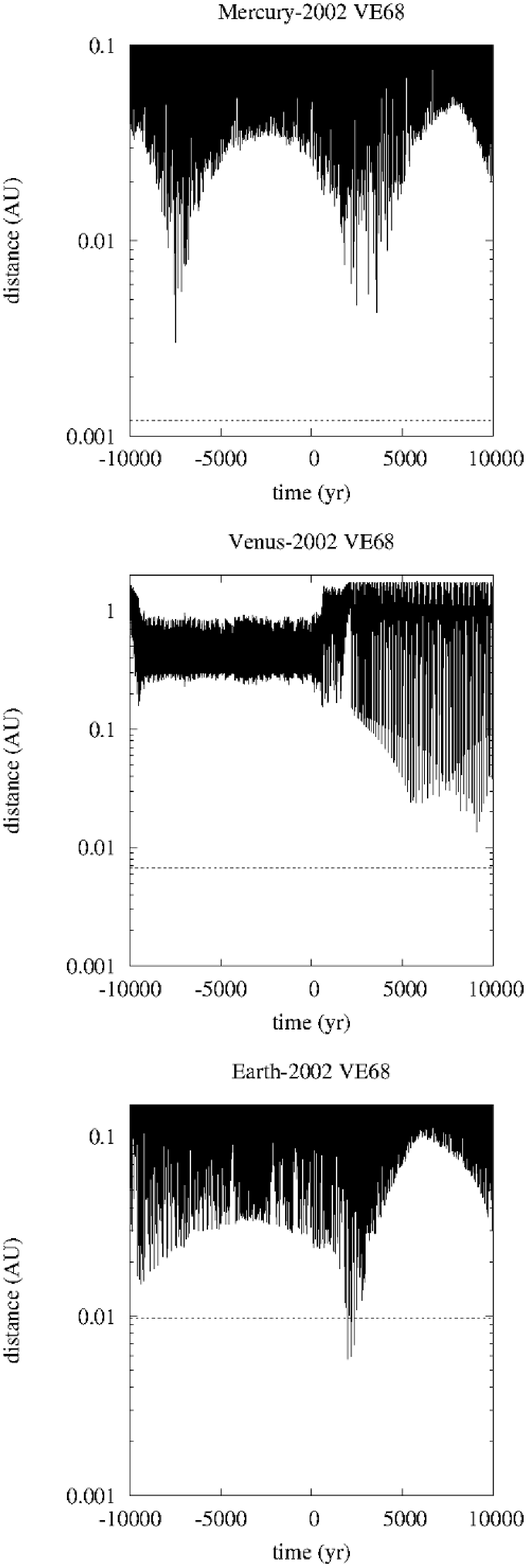,height=15cm,width=8cm,angle=0}
          }}
        \caption{The distance of 2002 VE$_{68}$ from Mercury (top panel), Venus (middle panel) and the Earth 
                 (bottom panel) from Model 2. The middle panel is equivalent to Fig. 3 in Mikkola et al. 
                 (2004). The bottom panel is equivalent to Fig. 4 in Mikkola et al. (2004).
                }
        \label{distance2}
     \end{figure}

%
%
        The inclusion of the Moon in the calculations has a significant impact on the evolution of the 
        distances in Fig. \ref{distance2}. Now encounters as close as 0.0025 AU (not shown) are possible for 
        both Mercury and the Earth so we may say that the resonant behaviour of 2002 VE$_{68}$ is controlled 
        by both the Earth and Mercury. On the other hand, relatively close encounters with Venus are only 
        observed prior to the quasi-satellite phase (multiple encounters under 0.03 AU) and after leaving the 
        quasi-satellite dynamical state (0.013 AU).
%
%
%
%

     \subsection{Model 3}
\begin{figure}
        \centerline{\hbox{
         \epsfig{figure=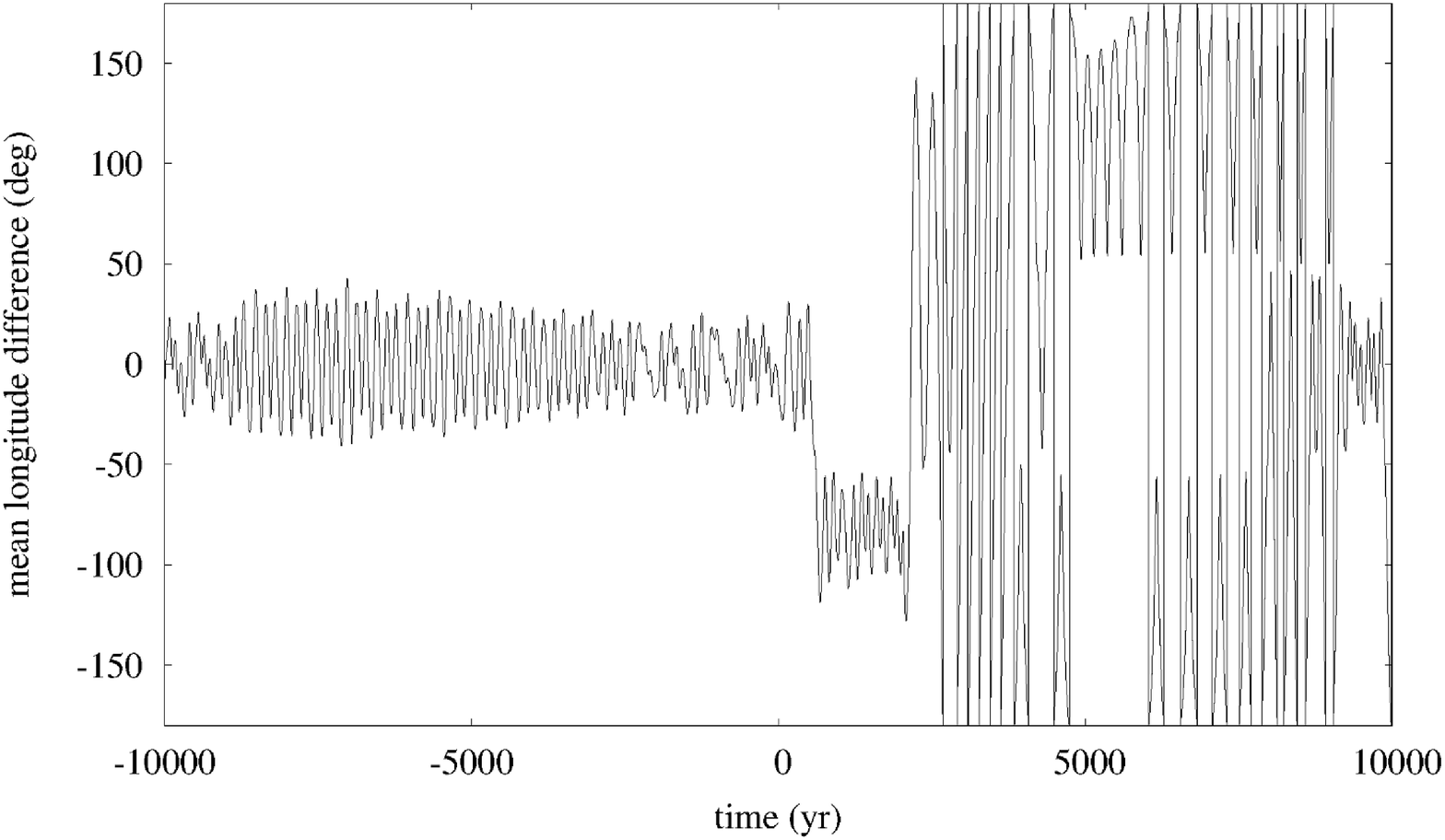,height=6cm,width=8cm,angle=0}
          }}
        \caption{The mean longitude difference of 2002 VE$_{68}$ and Venus in the time interval (-10,000, 10,
                 000) yr from Model 3, our most realistic model. In the time interval -7,000 to +1,000 yr from 
                 now the evolution of the mean longitude difference is very similar to that from Models 1 and 
                 2 (see Figs. \ref{mld1} and \ref{mld2}) but, after leaving the quasi-satellite dynamical 
                 state, the asteroid follows a very complex evolution with multiple transitions between
                 the various co-orbital resonant states. 
                }
        \label{mld3}
     \end{figure}

        The JPL Small-Body Database indicates that the three largest asteroids, (1) Ceres, (2) Pallas and (4) 
        Vesta, are minor perturbers of 2002 VE$_{68}$ and in our third and more realistic model we include the
        perturbations of eight planets, the Moon, (1) Ceres, (2) Pallas and (4) Vesta on the asteroid for the 
        nominal orbit in Table \ref{elements}. Figure \ref{mld3} shows that the evolution of the mean 
        longitude difference in the time interval -7,000 to +1,000 yr is very similar to those from Models 1 
        and 2 (see Figs. \ref{mld1} and \ref{mld2}) confirming again that 2002 VE$_{68}$ has been a 
        quasi-satellite of Venus for a period of time but that it will soon leave the quasi-satellite phase 
        for the L$_5$ Lagrangian point. The behavior of the mean longitude difference in that time range is
        consistent across models and differences only appear outside that time interval. The mean longitude of 
        the asteroid currently librates around the value of the mean longitude of Venus with an amplitude of 
        40$^{\circ}$-60$^{\circ}$ and an average period of about 150 yr. 
        \hfil\par
%
%
\begin{figure*}
        \centerline{\hbox{
         \epsfig{figure=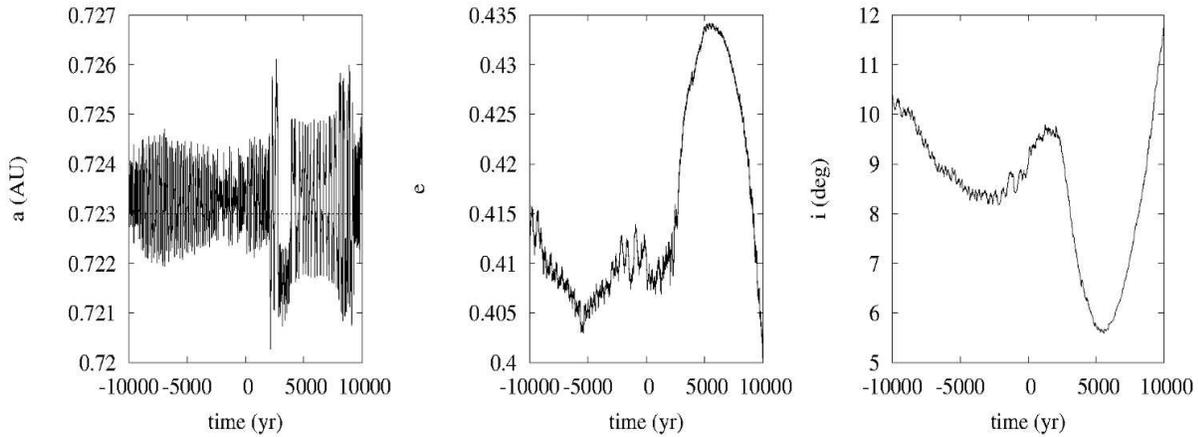,height=6cm,width=16cm,angle=0}
          }}
        \caption{Orbital elements of 2002 VE$_{68}$ from Model 3.
                }
        \label{orbele3}
     \end{figure*}

\begin{figure}
        \centerline{\hbox{
         \epsfig{figure=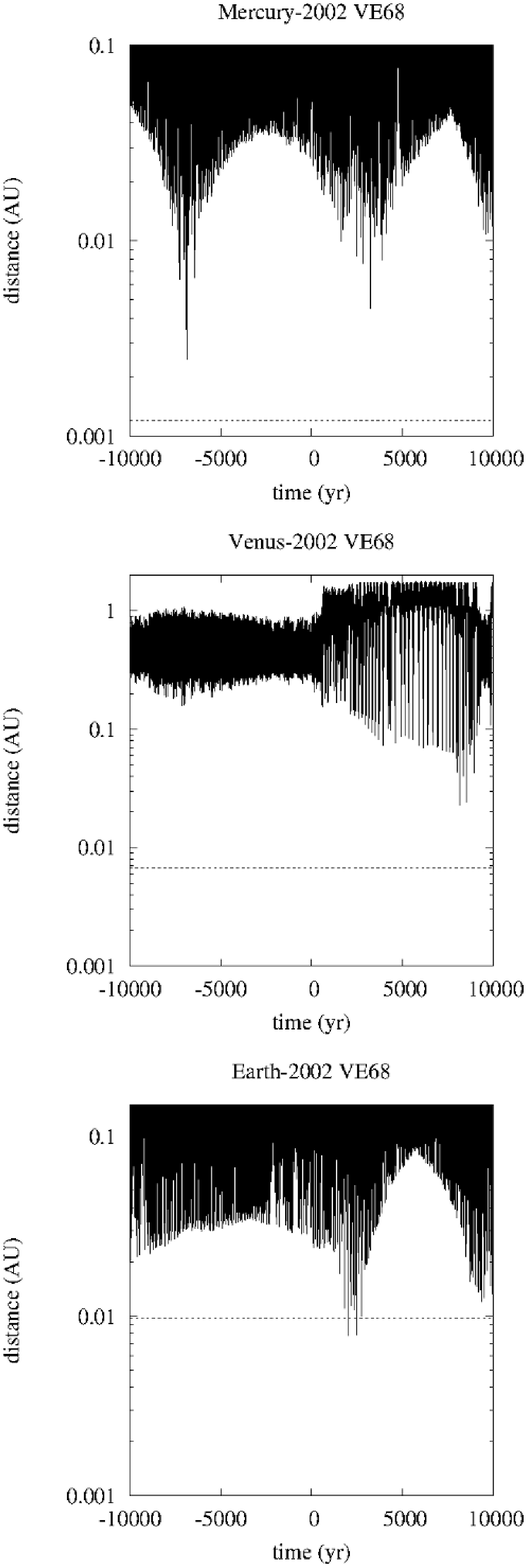,height=15cm,width=8cm,angle=0}
          }}
        \caption{The distance of 2002 VE$_{68}$ from Mercury (top panel), Venus (middle panel) and the Earth 
                 (bottom panel) from Model 3. The middle panel is equivalent to Fig. 3 in Mikkola et al. 
                 (2004). The bottom panel is equivalent to Fig. 4 in Mikkola et al. (2004).
                }
        \label{distance3}
     \end{figure}

%
%
        The orbital evolution of the asteroid after leaving the quasi-satellite phase is significantly more 
        complex than in previous models with multiple transitions between the various co-orbital resonant 
        states. The evolution of the orbital elements in Fig. \ref{orbele3} is quite similar to that in Fig. 
        \ref{orbele2} for Model 2. Figure \ref{distance3} is consistent with Fig. \ref{distance2} for Model 2 
        although close encounters with Venus after the end of the quasi-satellite phase tend to be more 
        distant; however, encounters as close as 0.002 AU are observed about 18,000 yr into the future. As in 
        the previous case, encounters with Mercury as close as 0.0025 AU are possible. In contrast, the 
        closest encounters with the Earth are now at 0.006 AU but still well within its Hill sphere. 
%
%
%
%
%
%
%
%

  \section{Discussion} 
\begin{figure}
        \centerline{\hbox{
         \epsfig{figure=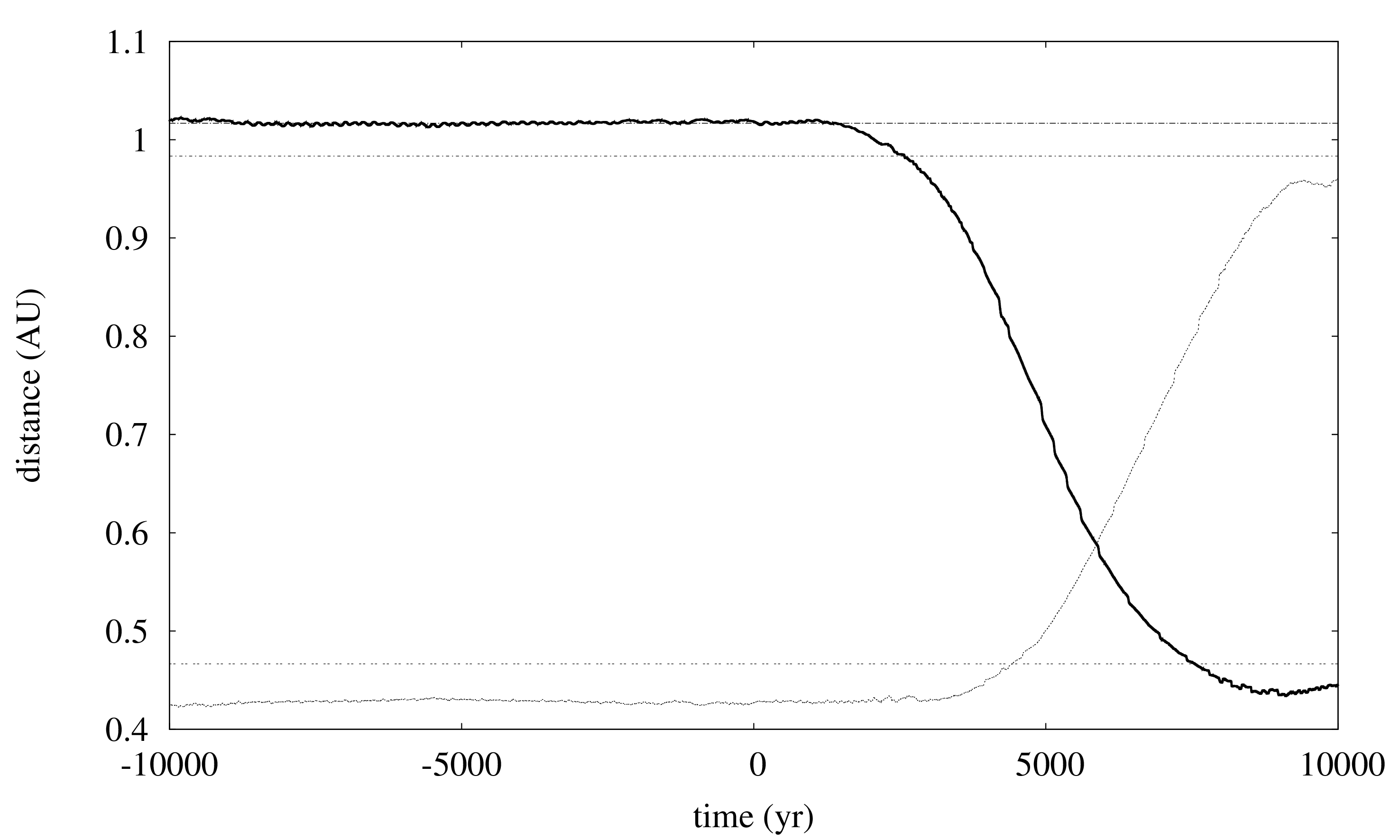,height=5cm,width=8cm,angle=0}
          }}
        \caption{The distance to the descending (thick line) and ascending nodes (dotted line) of 2002 
                 VE$_{68}$. Earth's aphelion and perihelion and Mercury's aphelion distances are also shown. 
                 The distance of the descending node coincides almost perfectly with Earth's aphelion during 
                 the entire quasi-satellite phase. This figure is equivalent to Fig. 5 in Mikkola et al. 
                 (2004).
                }
        \label{node}
     \end{figure}

     The asteroid 2002 VE$_{68}$ follows a rather eccentric orbit ($e\approx0.4$). In general, minor body 
     trajectories crossing the paths of one or more planets are rapidly destabilized by scatterings resulting 
     from close planetary approaches. The lifetime of the orbits of such objects can be relatively short. But
     this is only true if the orbital inclination is small, 2002 VE$_{68}$ moves in a relatively highly 
     inclined orbit ($i \approx 9^{\circ}$). In the Solar System and for a minor body moving in an inclined 
     orbit, close encounters with major planets are only possible in the vicinity of the nodes. The distance 
     between the Sun and the nodes is given by $r = a (1 - e^2) / (1 \pm e \cos \omega)$, where the "+" sign 
     is for the ascending node and the "-" sign is for the descending node. Figure \ref{node} shows the 
     evolution of the distance to the nodes of 2002 VE$_{68}$ along the studied time range. During the 
     quasi-satellite phase the distance to the descending node of the object remains remarkably close to the 
     value of Earth's aphelion and the distance to the ascending node is also relatively close to Mercury's 
     aphelion. In this way, the gravitational perturbations from the Earth (mainly) and Mercury are most 
     effective in keeping the asteroid at a safe distance from Venus. The object approaches the Earth in the 
     vicinity of its descending node in November every 8th year. In fact, 2002 VE$_{68}$ orbits the Sun in a 
     near 8:13 resonance with the Earth and, currently, its orbit appears to be stabilized by close 
     encounters to the Earth every 8 years. 2002 VE$_{68}$ has a period of 0.6156 yr that is almost 8/13 
     (0.6154) so the Earth completes 8 orbits around the Sun in the same amount of time the asteroid 
     completes 13. But this is not all, 2002 VE$_{68}$ is also moving in a near 9:23 resonance with Mercury. 
     2002 VE$_{68}$ exhibits resonant (or near resonant) behaviour with Mercury, Venus and the Earth. 
     \hfil\par
%
%
\begin{figure}
        \centerline{\hbox{
         \epsfig{figure=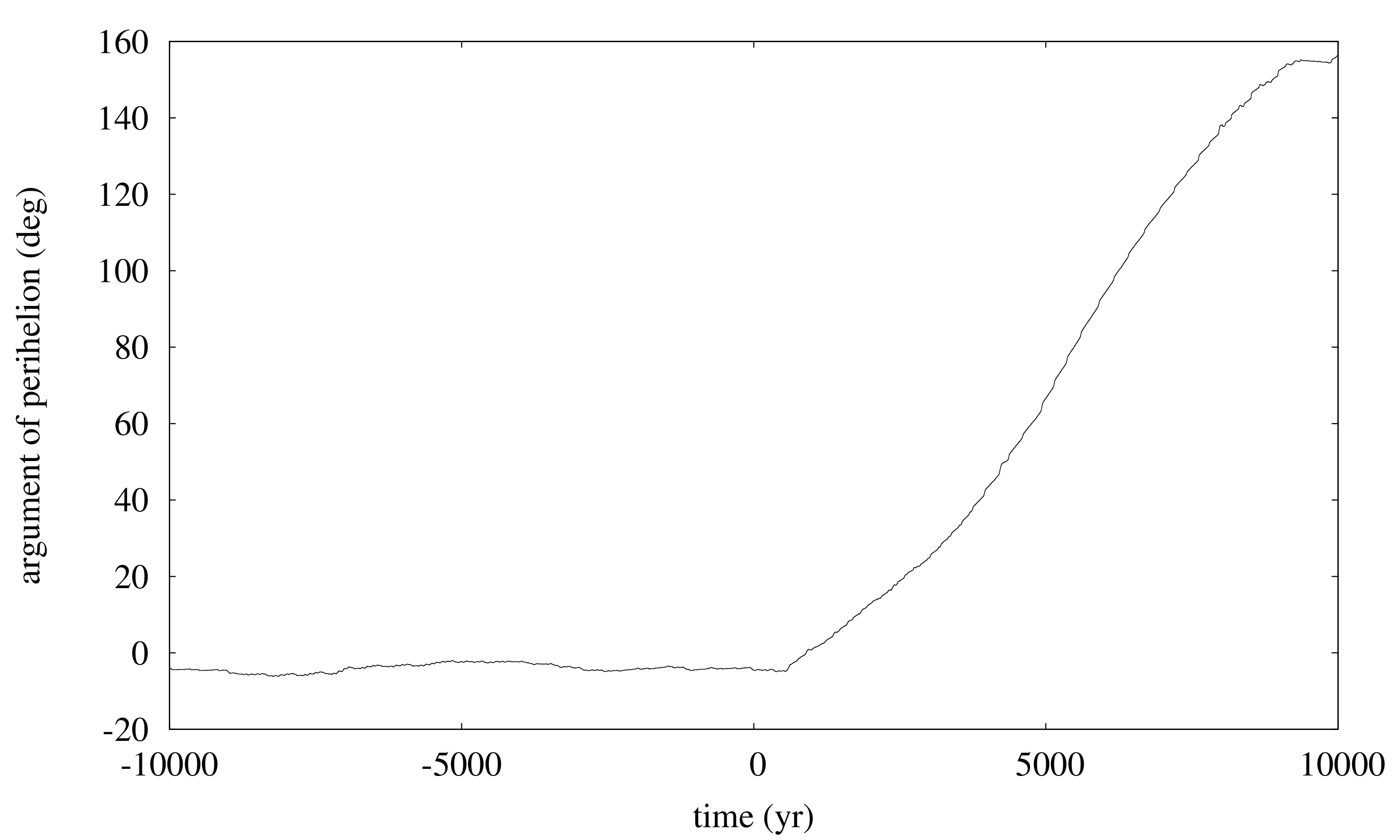,height=5cm,width=8cm,angle=0}
          }}
        \caption{The argument of perihelion of 2002 VE$_{68}$. Its value decreases during the quasi-satellite 
                 phase at a rate of $\dot{\omega} = -0.0005^{\circ}$/yr. 
                }
        \label{w}
     \end{figure}

%
%
     On the other hand and after the object leaves its quasi-satellite path, it undergoes multiple transitions 
     between resonant states (see Fig. \ref{mld3}). This significant complexity is the result of having high 
     eccentricity and inclination, in this case compound orbits are possible (Namouni 1999; Namouni, Christou 
     \& Murray 1999). When moving in a 1:1 mean motion resonance, changes in the values of the semi-major 
     axis, the eccentricity and the inclination are small compared to the variation of the argument of 
     perihelion. Transfers between quasi-satellite, horseshoe and tadpole orbits are the result of the 
     libration of the nodes (Wiegert, Innanen \& Mikkola 1998). The argument of perihelion of 2002 VE$_{68}$ 
     is displayed in Fig. \ref{w}. During the quasi-satellite phase, its value decreases at a rate of 
     $\dot{\omega} = -0.0005^{\circ}$/yr. This secular change in the value of the argument of perihelion was 
     predicted by Namouni (1999) on theoretical grounds and can be used to precisely track transitions between 
     the quasi-satellite state and any other. Based on the precession of the argument of perihelion criterion 
     and for Model 3, the object enters the quasi-satellite phase at about -14,000 yr and it leaves the phase 
     555 yr from now. 
     \hfil\par
%
%
\begin{figure}
        \centerline{\hbox{
         \epsfig{figure=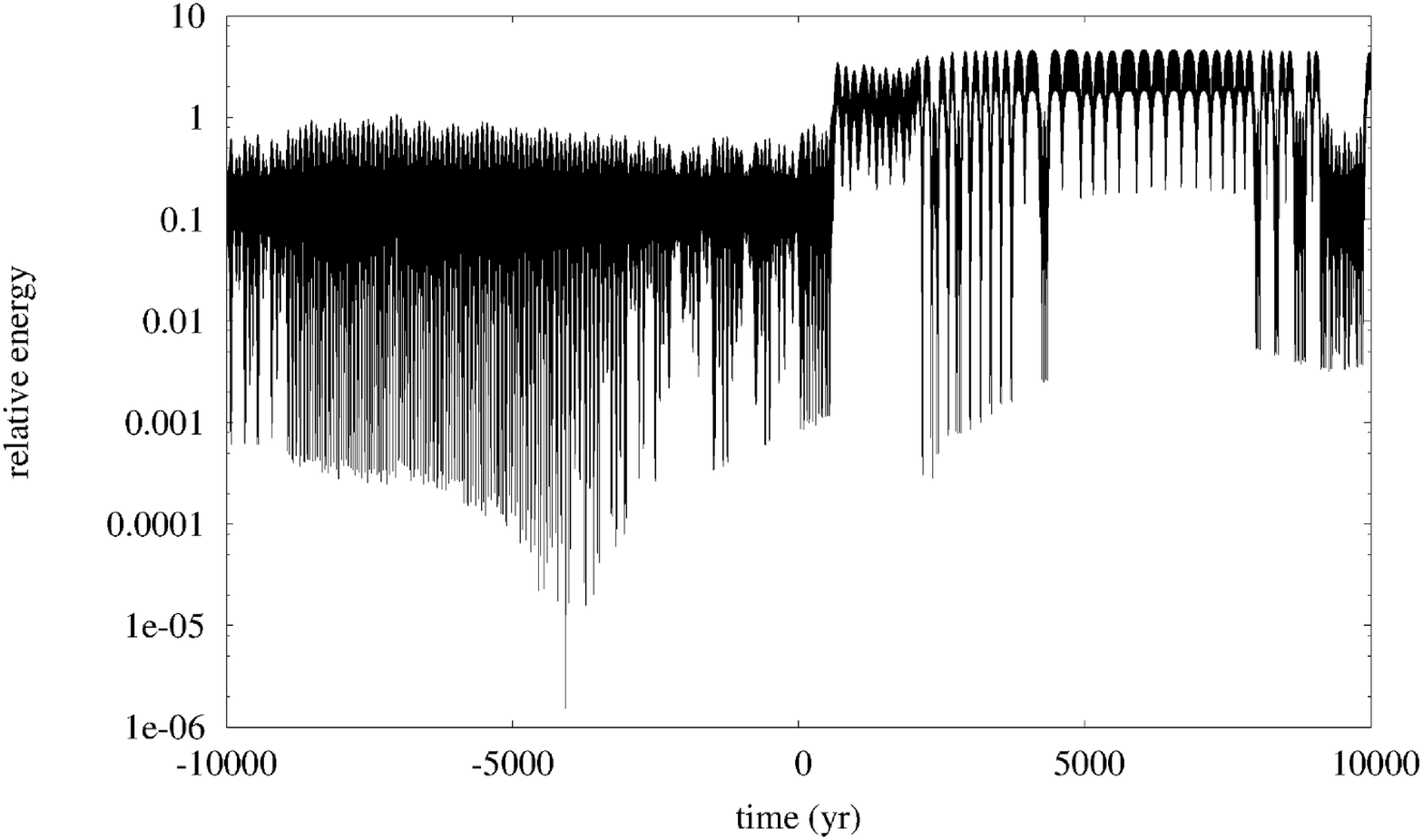,height=5cm,width=8cm,angle=0}
          }}
        \caption{Total energy (specific orbital energy) of 2002 VE$_{68}$ relative to Venus. The 
                 quasi-satellite state, even if not bound (energy $>$ 0), is significantly less energetic than 
                 the other resonant states (Trojan or other). 
                }
        \label{energy}
     \end{figure}

%
%
     Regarding the energy balance relative to the host planet involved in the various resonant transitions, in 
     Fig. \ref{energy} we show the total energy (specific orbital energy) of 2002 VE$_{68}$ relative to Venus. 
     The quasi-satellite dynamical state, even if not bound (the total relative energy is still $>$ 0), is 
     significantly less energetic than the other resonant states (Trojan or any other). If, during the 
     quasi-satellite phase the object suffers any significant decceleration as a result of, for example, drag 
     or a distant interaction with another body (perhaps a pre-existing natural satellite) or ejection of one
     of the components in a binary asteroid, the object (or one of the components of a hypothetical binary 
     system) may be permanently trapped in a retrograde orbit around the host planet. With no natural 
     satellites, this scenario is unlikely to work for Venus or Mercury (unless the incoming asteroid is a
     binary) but it may be valid in other cases. 
     \hfil\par
%
%
%
%
     The relatively long-term quasi-satellite dynamical state of 2002 VE$_{68}$ has been further confirmed 
     using control orbits obtained from the nominal orbit by varying the orbital elements within the error
     range of the observed object (see Table \ref{elements}). All the control orbits exhibit consistent
     behaviour in the time range -7,000 to +1,000 yr. The orbital behaviour on longer time-scales, even if not
     coincident, can also be discussed but in probabilistic terms. For example, 100\% of the control orbits go 
     from the $L_4$ Lagrangian point into the quasi-satellite phase but that transition takes place in the 
     time frame 8,000-7,000 yr into the past for 70\% of the control orbits with the remaining 30\% 
     experiencing the transition earlier than 8,000 yr ago. Prior to the $L_4$ Trojan state, 60\% of control 
     calculations were following horseshoe orbits, 20\% were in the quasi-satellite state, 10\% were in the 
     $L_5$ Trojan state and the remaining 10\% were following a classical, non-resonant passing orbit. So 
     very likely and prior to the quasi-satellite phase, the object was already co-orbital with Venus and 
     about 7,500 yr ago it became a quasi-satellite after a close encounter with the Earth.
     \hfil\par
     Regarding its NEO status, our calculations indicate that an actual collision with the Earth during the 
     next 10,000 yr is highly unlikely but a relatively close approach at 0.03782 AU on November 4, 2018, 
     12:41 UT will take place (the MPC\footnote{http://www.minorplanetcenter.net/iau/lists/PHACloseApp.html} 
     quotes 0.03764 AU on November 4.90, 2018). Encounters that close to the Earth occur regularly and they 
     have a periodicity of 8 years. In fact, 2002 VE$_{68}$ was discovered during one of these close 
     approaches, November 11, 2002. This periodicity is explained as a result of the near resonant behaviour 
     with the Earth (see above). Within the next 100 yr, the closest approach will take place on November 12, 
     2106 at 0.02482 AU. No encounters closer than 0.023 AU will be recorded within the next 1,000 yr.
     \hfil\par
     As for the apparently battered surface of 2002 VE$_{68}$ we cannot avoid to speculate whether the 
     relatively strong tidal forces generated during its frequent close fly-bys with the inner planets may 
     have caused the object to evolve into an aggregate of shattered pieces. Also high velocity encounters 
     with other minor bodies may have played a role as the object moves in a rather eccentric orbit in the 
     inner solar system.

  \section{Comparison with Mikkola et al. (2004)}
     Mikkola et al. (2004) identified 2002 VE$_{68}$ as the first {\it bona fide} quasi-satellite. They found 
     that it has remained in the quasi-satellite dynamical state for about 7,000 years and it will remain in 
     that phase for another 500 years. Then it will move into a temporary tadpole orbit around Venus L$_{5}$ 
     Lagrangian point becoming a Trojan asteroid for about 700 years to later transfer to the L$_{4}$ 
     Lagrangian point. In their work, they pointed out that due to its large eccentricity, 2002 VE$_{68}$ 
     experiences frequent close approaches to the Earth as the asteroid descending node stays close to the 
     Earth's orbit. They concluded that although the Earth plays a major role in the orbital evolution of 2002 
     VE$_{68}$, all close encounters are well outside its Hill sphere. In their calculations, a version of the 
     second-order Wisdom-Holman symplectic map is used (Wisdom \& Holman 1991) with a time-step of 0.1 d. 
     Mikkola et al. (2004) do not provide many details on their physical model: the actual number of planets 
     (or any other objects) included in their calculations is not given, the initial conditions are not 
     clearly stated, energy or angular momentum conservation are not discussed. However, our results are in 
     general compatible and consistent with theirs although our close encounters with the Earth and Mercury
     in Models 2 and 3 are significantly closer than those reported by Mikkola et al. (2004), likely as a
     result of using a separate body for the Moon in our calculations. 
     \hfil\par
     We confirm the Venus quasi-satellite nature of 2002 VE$_{68}$, also that it will leave its unusual 
     dynamic status in a relatively short time-scale (about 500 yr) and it got into its current state after a 
     close encounter with the Earth about 7,000 yr ago, although the actual time-scale for this event is less 
     constrained in our calculations (7,000-14,000 yr ago). We agree that the Earth has a dominant role on 
     the orbital evolution of 2002 VE$_{68}$ but Mercury is also a secondary player and due to the close 
     encounters with the Earth-Moon system, the role of the Moon cannot be neglected. We also agree that the 
     effect of the precession of the argument of perihelion on the upper nodal point of the object is the 
     actual mechanism engaging or disengaging the quasi-satellite phase: the distance from the Sun to the 
     upper nodal point coincides quite well with the value of Earth's aphelion. They found that the e-folding
     time during the quasi-satellite phase was nearly 300 yr; our calculations give $\sim$200 yr during the
     same period. It is also obvious from our calculations that a relatively strong gravitational interaction 
     with the Earth injected the asteroid into its present orbit and eventually will be responsible for its 
     future transition to a different co-orbital resonant state. The consistency between results from 
     different integrators and different models shows that the results are solid and statistically robust.
  
  \section{Conclusions} 
     2002 VE$_{68}$, a remarkable NEO, was discovered by B. Skiff in 2002 (Griesser et al. 2002) and 
     subsequently identified as a quasi-satellite of Venus (Mikkola et al. 2004). This paper revisits the 
     dynamical status of 2002 VE$_{68}$, numerically integrating its trajectory using updated ephemerides and 
     analyzing the results. We studied the orbit of 2002 VE$_{68}$ using different models and found good 
     agreement between them on short time-scales. We can summarize the results of our investigation as 
     follows:

       a) We confirm the Venus quasi-satellite nature of 2002 VE$_{68}$ announced by Mikkola et al. (2004).  

       b) We confirm that 2002 VE$_{68}$ will leave its unusual dynamic status in a relatively short 
          time-scale (about 500 yr). 

       c) We confirm that 2002 VE$_{68}$ got into its actual state after a close encounter with the Earth 
          about 7,000-14,000 yr ago.  

       d) Close approaches are possible both at perihelion (with Mercury) and aphelion (with the Earth). 
          Earth's are more important. 

       e) The influence of the Moon on the dynamics of 2002 VE$_{68}$ is not negligible as very close 
          encounters with the Earth are possible. 

       f) 2002 VE$_{68}$ exhibits resonant (or near resonant) behaviour with Mercury, Venus and the Earth.

       g) There is no danger of impact with the Earth, Venus or Mercury in the near future. Relatively
          close encounters with our planet have a periodicity of 8 years. The next close approach to the Earth 
          will take place on November 4, 2018 at 0.038 AU. 

     Currently, the Earth has five known objects regarded as quasi-satellites: 3753 Cruithne (Wiegert, Innanen 
     \& Mikkola 1997), 2003 YN$_{107}$ (Connors et al. 2004; Brasser et al. 2004), (164207) 2004 GU$_{9}$ 
     (Connors et al. 2004; Brasser et al. 2004; Wiegert et al. 2005; Mikkola et al. 2006; Wajer 2010), 2006 
     FV$_{35}$ (Mikkola et al. 2006; Stacey \& Connors 2009; Wajer 2010) and 2010 SO$_{16}$ (Christou \& Asher 
     2011). 2002 AA$_{29}$ will become a quasi-satellite of the Earth in the future (Connors et al. 2002). 
     Venus has one, 2002VE$_{68}$ (Mikkola et al. 2004). Jupiter has four known quasi-satellites (Kinoshita \& 
     Nakai 2007): 2001 QQ$_{199}$, 2004 AE$_9$, P/2002 AR$_2$ LINEAR and P/2003 WC$_7$ LINEAR-CATALINA but new 
     co-orbitals have recently been identified (Wajer \& Kr\'olikowska 2012). A few hundred Main Belt 
     asteroids appear to be co-orbital (quasi-satellites in some cases) with (1) Ceres and (4) Vesta (Christou 
     2000; Christou \& Wiegert 2012). Clearly the more objects identified in the quasi-satellite phase the 
     better our understanding on their stability will be. Recognizing a variety of objects in the 
     quasi-satellite state under different dynamical environments can only improve our knowledge on the 
     overall processes that lead to the transformation of passing orbits into co-orbital ones and viceversa. 

  \section*{Acknowledgements}
     The authors would like to thank S. Aarseth for providing the codes used in this research and for comments
     on the manuscript. We also thank the referee for her/his prompt review. This work was partially supported 
     by the Spanish 'Comunidad de Madrid' under grant CAM S2009/ESP-1496 (Din\'amica Estelar y Sistemas 
     Planetarios). We thank Dr. Mar\'{\i}a Jos\'e Fern\'andez-Figueroa, Dr. Manuel Rego Fern\'andez and the 
     Department of Astrophysics of Universidad Complutense de Madrid for providing excellent computing 
     facilities. Most of the calculations and part of the data analysis were completed on the 'Servidor 
     Central de C\'alculo' of the Universidad Complutense of Madrid and we thank the computing staff (Santiago 
     Cano Als\'ua) for their help during this stage. In preparation of this paper, we made use of the NASA 
     Astrophysics Data System and the ASTRO-PH e-print server.

\end{document}